\documentstyle[12pt]{article}

\skewchar\fivmi='177
\skewchar\sixmi='177
\skewchar\sevmi='177
\skewchar\egtmi='177
\skewchar\ninmi='177
\skewchar\tenmi='177
\skewchar\elvmi='177
\skewchar\twlmi='177
\skewchar\frtnmi='177
\skewchar\svtnmi='177
\skewchar\twtymi='177
\def\@magscale#1{ scaled \magstep #1}
\font\twfvmi  = ammi10   \@magscale5 
\skewchar\twfvmi='177
\skewchar\fivsy='60
\skewchar\sixsy='60
\skewchar\sevsy='60
\skewchar\egtsy='60
\skewchar\ninsy='60
\skewchar\tensy='60
\skewchar\elvsy='60
\skewchar\twlsy='60
\skewchar\frtnsy='60
\skewchar\svtnsy='60
\skewchar\twtysy='60
\font\twfvsy  = amsy10   \@magscale5 
\skewchar\twfvsy='60

\catcode`@=11
\def\un#1{\relax\ifmmode\@@underline#1\else
        $\@@underline{\hbox{#1}}$\relax\fi}
\catcode`@=12

\let\du=\d                      
\let\um=\H                      

\def\a{\alpha}
\def\b{\beta}

\def\d{\delta}
\def\e{\epsilon}

\def\g{\gamma}

\def\l{\lambda}
\def\m{\mu}
\def\n{\nu}

\def\r{\rho}
\def\s{\sigma}
\def\t{\tau}

\def\L{\Lambda}

\font\sc=font005                        
\def\Sc#1{{\hbox{\sc #1}}}      
\font\ooo=circle10                      
\font\ro=manfnt                         
\def\kcl{{\hbox{\ro 6}}}                
\def\kcr{{\hbox{\ro 7}}}                
\def\ktl{{\hbox{\ro \char'134}}}        
\def\ktr{{\hbox{\ro \char'135}}}        
\def\kbl{{\hbox{\ro \char'136}}}        
\def\kbr{{\hbox{\ro \char'137}}}        

\def\ip{{=\!\!\! \mid}}                                    
\def\bo{{\raise.15ex\hbox{\large$\Box$}}}               
\def\pr{\prod}                                          
\def\TH{{\raise.2ex\hbox{$\displaystyle \bigodot$}\mskip-4.7mu \llap H \;}}
\def\face{{\raise.2ex\hbox{$\displaystyle \bigodot$}\mskip-2.2mu \llap {$\ddot
        \smile$}}}                                      

\def\sp#1{{}^{#1}}                              
\def\Tilde#1{{\widetilde{#1}}\hskip 0.03in}                     
\def\Hat#1{\widehat{#1}}                        
\def\Bar#1{\overline{#1}}                       
\def\leftrightarrowfill{$\mathsurround=0pt \mathord\leftarrow \mkern-6mu
        \cleaders\hbox{$\mkern-2mu \mathord- \mkern-2mu$}\hfill
        \mkern-6mu \mathord\rightarrow$}
\def\dvec#1{\vbox{\ialign{##\crcr
        \leftrightarrowfill\crcr\noalign{\kern-1pt\nointerlineskip}
        $\hfil\displaystyle{#1}\hfil$\crcr}}}           
\def\dt#1{{\buildrel {\hbox{\LARGE .}} \over {#1}}}     

\def\frac#1#2{{\textstyle{#1\over\vphantom2\smash{\raise.20ex
        \hbox{$\scriptstyle{#2}$}}}}}                   
\def\ha{\frac12}                                        
\def\sfrac#1#2{{\vphantom1\smash{\lower.5ex\hbox{\small$#1$}}\over
        \vphantom1\smash{\raise.4ex\hbox{\small$#2$}}}} 
\def\bfrac#1#2{{\vphantom1\smash{\lower.5ex\hbox{$#1$}}\over
        \vphantom1\smash{\raise.3ex\hbox{$#2$}}}}       
\def\afrac#1#2{{\vphantom1\smash{\lower.5ex\hbox{$#1$}}\over#2}}    

\newskip\humongous \humongous=0pt plus 1000pt minus 1000pt
\def\caja{\mathsurround=0pt}
\def\eqalign#1{\,\vcenter{\openup2\jot \caja
        \ialign{\strut \hfil$\displaystyle{##}$&$
        \displaystyle{{}##}$\hfil\crcr#1\crcr}}\,}
\newif\ifdtup

\def\ref#1{$\sp{#1)}$}

\parskip=\medskipamount                 
\thicklines                         

\def\oldheadpic{                                
        \setlength{\unitlength}{.4mm}
        \thinlines
        \par
        \begin{picture}(349,16)
        \put(325,16){\line(1,0){4}}
        \put(330,16){\line(1,0){4}}
        \put(340,16){\line(1,0){4}}
        \put(335,0){\line(1,0){4}}
        \put(340,0){\line(1,0){4}}
        \put(345,0){\line(1,0){4}}
        \put(329,0){\line(0,1){16}}
        \put(330,0){\line(0,1){16}}
        \put(339,0){\line(0,1){16}}
        \put(340,0){\line(0,1){16}}
        \put(344,0){\line(0,1){16}}
        \put(345,0){\line(0,1){16}}
        \put(329,16){\oval(8,32)[bl]}
        \put(330,16){\oval(8,32)[br]}
        \put(339,0){\oval(8,32)[tl]}
        \put(345,0){\oval(8,32)[tr]}
        \end{picture}
        \par
        \thicklines
        \vskip.2in}
\def\oldtitle#1#2#3#4{\oldheadpic\begin{center}\vglue.5in{\large\bf #1}\\[.6in]
        {#2}\\[.1in] {\it Department of Physics and Astronomy}\\
        {\it University of Maryland, College Park, MD 20742}\\[.6in]
        Physics Publication \#{#3}\\ {#4}\\[1.5in] {\bf Abstract}\\[.1in]
        \end{center} \begin{quotation}}                 
\def\oldTitle#1#2#3#4#5#6#7{\oldheadpic\begin{center} \vglue .4in
        {\large\bf #1}\\[.4in]
        {#2}\\[.1in] {\it Department of Physics and Astronomy}\\
        {\it University of Maryland, College Park, MD 20742}\\[.1in]
        {#3}\\[.1in] {\it {#4}}\\ {\it {#5}}\\[.4in]
        Physics Publication \#{#6}\\ {#7}\\[.5in] {\bf Abstract}\\[.1in]
        \end{center} \begin{quotation}}                 
\def\border{                                            
        \setlength{\unitlength}{1mm}
        \newcount\xco
        \newcount\yco
        \xco=-24
        \yco=12
        \begin{picture}(140,0)
        \put(\xco,\yco){$\ktl$}
        \advance\yco by-1
        {\loop
        \put(\xco,\yco){$\kcl$}
        \advance\yco by-2
        \ifnum\yco>-240
        \repeat
        \put(\xco,\yco){$\kbl$}}
        \xco=158
        \yco=12
        \put(\xco,\yco){$\ktr$}
        \advance\yco by-1
        {\loop
        \put(\xco,\yco){$\kcr$}
        \advance\yco by-2
        \ifnum\yco>-240
        \repeat
        \put(\xco,\yco){$\kbr$}}
        \put(-20,11){\tiny University of Maryland Elementary Particle
Physics University of Maryland Elementary Particle Physics University of
Maryland Elementary Particle Physics}
        \put(-20,-241.5){\tiny University of Maryland Elementary
Particle Physics University of Maryland Elementary Particle Physics
University of Maryland Elementary Particle Physics}
        \end{picture}
        \par\vskip-8mm}
\def\bordero{                                           
        \setlength{\unitlength}{1mm}
        \newcount\xco
        \newcount\yco
        \xco=-24
        \yco=12
        \begin{picture}(140,0)
        \put(\xco,\yco){$\ktl$}
        \advance\yco by-1
        {\loop
        \put(\xco,\yco){$\kcl$}
        \advance\yco by-2
        \ifnum\yco>-240
        \repeat
        \put(\xco,\yco){$\kbl$}}
        \xco=158
        \yco=12
        \put(\xco,\yco){$\ktr$}
        \advance\yco by-1
        {\loop
        \put(\xco,\yco){$\kcr$}
        \advance\yco by-2
        \ifnum\yco>-240
        \repeat
        \put(\xco,\yco){$\kbr$}}
        \put(-20,12){\ooo
bacdefghidfghghdhededbihdgdfdfhhdheidhdhebaaahjhhdahba
hgdedge
   hgfdiehhgdigicba}
        \put(-20,-241.5){\ooo
ababaighefdbfghgeahgdfgafagihdidihiidhiagfedhadbfd
ecdcdfa
   gdcbhaddhbgfchbgfdacfediacbabab}
        \end{picture}
        \par\vskip-8mm}
\def\headpic{                                           
        \indent
        \setlength{\unitlength}{.4mm}
        \thinlines
        \par
        \begin{picture}(29,16)
        \put(165,16){\line(1,0){4}}
        \put(170,16){\line(1,0){4}}
        \put(180,16){\line(1,0){4}}
        \put(175,0){\line(1,0){4}}
        \put(180,0){\line(1,0){4}}
        \put(185,0){\line(1,0){4}}
        \put(169,0){\line(0,1){16}}
        \put(170,0){\line(0,1){16}}
        \put(179,0){\line(0,1){16}}
        \put(180,0){\line(0,1){16}}
        \put(184,0){\line(0,1){16}}
        \put(185,0){\line(0,1){16}}
        \put(169,16){\oval(8,32)[bl]}
        \put(170,16){\oval(8,32)[br]}
        \put(179,0){\oval(8,32)[tl]}
        \put(185,0){\oval(8,32)[tr]}
        \end{picture}
        \par\vskip-6.5mm
        \thicklines}
\def\title#1#2#3#4{\border\headpic {\hbox to\hsize{#4 \hfill UMDEPP #3}}\par
        \begin{center} \vglue .5in {\large\bf #1}\\[.6in]
        {#2}\\[.1in] {\it Department of Physics and Astronomy}\\
        {\it University of Maryland, College Park, MD 20742}\\[1.5in]
        {\bf Abstract}\\[.1in] \end{center} \begin{quotation}}  
\def\Title#1#2#3#4#5#6#7{\border\headpic
        {\hbox to\hsize{#7 \hfill UMDEPP #6}}\par
        \begin{center} \vglue .4in {\large\bf #1}\\[.4in]
        {#2}\\[.1in] {\it Department of Physics and Astronomy}\\
        {\it University of Maryland, College Park, MD 20742}\\[.1in]
        {#3}\\[.1in] {\it {#4}}\\ {\it {#5}}\\[.5in] {\bf Abstract}\\[.1in]
        \end{center} \begin{quotation}}                 
\def\endtitle{\end{quotation}\newpage}                  

\def\sect#1{\bigskip\medskip \goodbreak \noindent{\bf {#1}} \nobreak \medskip}
\def\refs{\sect{References} \footnotesize \frenchspacing \parskip=0pt}
\def\Item{\par\hang\textindent}

\begin{document}

\def\scst{\scriptstyle}
\def\itrema{$\ddot{\scriptstyle 1}$}
\def\Bo{\bo{\hskip 0.03in}}
\def\lrad#1{ \left( A {\buildrel\leftrightarrow\over D}_{#1} B\right) }
\def\derx{\partial_x} \def\dery{\partial_y} \def\dert{\partial_t}
\def\Vec#1{{\overrightarrow{#1}}}
\def\.{.$\,$}

\def\grg#1#2#3{Gen.~Rel.~Grav.~{\bf{#1}} (19{#2}) {#3} }

\def\pla#1#2#3{Phys.~Lett.~{\bf A{#1}} (19{#2}) {#3}}

\def\ula{{\underline a}} \def\ulb{{\underline b}} \def\ulc{{\underline c}}
\def\uld{{\underline d}} \def\ule{{\underline e}} \def\ulf{{\underline f}}
\def\ulg{{\underline g}} \def\ulm{{\underline m}}
\def\uln#1{\underline{#1}}
\def\ulp{{\underline p}} \def\ulq{{\underline q}} \def\ulr{{\underline r}}

\def\plpl{{+\!\!\!\!\!{\hskip 0.009in}{\raise -1.0pt\hbox{$_+$}}
{\hskip 0.0008in}}}

\def\mimi{{-\!\!\!\!\!{\hskip 0.009in}{\raise -1.0pt\hbox{$_-$}}
{\hskip 0.0008in}}}

\def\items#1{\\ \item{[#1]}}
\def\ul{\underline}
\def\un{\underline}
\def\-{{\hskip 1.5pt}\hbox{-}}

\def\kd#1#2{\d\du{#1}{#2}}
\def\fracmm#1#2{{{#1}\over{#2}}}
\def\footnotew#1{\footnote{\hsize=6.5in {#1}}}

\def\low#1{{\raise -3pt\hbox{${\hskip 1.0pt}\!_{#1}$}}}

\def\ip{{=\!\!\! \mid}}
\def\unb{{\underline {\bar n}}}
\def\upb{{\underline {\bar p}}}
\def\um{{\underline m}}
\def\up{{\underline p}}
\def\Phib{{\Bar \Phi}}
\def\Phit{{\tilde \Phi}}
\def\Phibt{{\tilde {\Bar \Phi}}}
\def\Db{{\Bar D}_{+}}
\def\gg{{\hbox{\sc g}}}
\def\nt{$~N=2$~}

\border\headpic {\hbox to\hsize{February 1993\hfill UMDEPP 93--144}}\par
\hfill {(Revised Version)}\\

\begin{center}
\vglue .25in

{\large\bf Self-Dual Supersymmetric Yang-Mills Theory} \\
{\large\bf Generates Witten's Topological Field Theory}$\,$\footnote{This
work is supported in part by NSF grant \# PHY-91-19746.} \\[.1in]

\baselineskip 10pt

\vskip 0.25in

Hitoshi NISHINO\footnote{E-Mail: Nishino@UMDHEP.bitnet} \\[.2in]
{\it Department of Physics} \\ [.015in]
{\it University of Maryland at College Park}\\ [.015in]
{\it College Park, MD 20742-4111, USA} \\[.1in]
and\\[.1in]
{\it Department of Physics and Astronomy} \\[.015in]
{\it Howard University} \\[.015in]
{\it Washington, D.C. 20059, USA} \\[.18in]

\vskip 1.0in

{\bf Abstract}\\[.1in]
\end{center}

\begin{quotation}

{}~~~We show that the recently constructed $~N=4$~ supersymmetric
self-dual Yang-Mills theory as the consistent background of
\hbox{$~N=2$} open superstring will generate Witten's topological field
theory in two-dimensions as a descendant theory after appropriate twisted
dimensional reduction/truncations.  We also show that this topological
field theory further generates supersymmetric Korteweg de Vries equations,
$~SL(n)\-$Toda theory and
$~W_\infty\-$gravity in the $~n\rightarrow\infty$~ limit.  Considering also
that this topological field theory is to generate other integrable
and topological systems, our results give supporting evidence for the
conjecture that the $~N=2$~ open superstring theory is
the underlying ``master theory'' of integrable and
topological systems, {\it via} the intermediate $~N=4$~ supersymmetric
self-dual Yang-Mills theory as its consistent target space-time background.

\endtitle

\def\doit#1#2{\ifcase#1\or#2\fi}
\def\[{\lfloor{\hskip 0.35pt}\!\!\!\lceil}
\def\]{\rfloor{\hskip 0.35pt}\!\!\!\rceil}
\def\delsl{{{\partial\!\!\! /}}}
\def\caldsl{{\calD\!\!\! /}}
\def\calO{{\cal O}}
\def\asym{({\scriptstyle 1\leftrightarrow \scriptstyle 2})}
\def\Lag{{\cal L}}
\def\du#1#2{_{#1}{}^{#2}}
\def\ud#1#2{^{#1}{}_{#2}}
\def\dud#1#2#3{_{#1}{}^{#2}{}_{#3}}
\def\udu#1#2#3{^{#1}{}_{#2}{}^{#3}}
\def\calD{{\cal D}}
\def\calM{{\cal M}}
\def\tildef{{\tilde f}}
\def\calDsl{{\calD\!\!\!\! /}}

\def\Hat#1{{#1}{\large\raise-0.02pt\hbox{$\!\hskip0.038in\!\!\!\hat{~}$}}}
\def\hati{{\hat{I}}}
\def\dt{$~D=10$~}
\def\alp{\alpha{\hskip 0.007in}'}
\def\oalp#1{\alp^{\hskip 0.007in {#1}}}
\def\naive{{{na${\scriptstyle 1}\!{\dot{}}\!{\dot{}}\,\,$ve}}}
\def\items#1{\vskip 0.05in\Item{[{#1}]}}
\def\item#1{\Item{#1}}

\def\pl#1#2#3{Phys.~Lett.~{\bf {#1}B} (19{#2}) #3}
\def\np#1#2#3{Nucl.~Phys.~{\bf B{#1}} (19{#2}) #3}
\def\prl#1#2#3{Phys.~Rev.~Lett.~{\bf #1} (19{#2}) #3}
\def\pr#1#2#3{Phys.~Rev.~{\bf D{#1}} (19{#2}) #3}
\def\cqg#1#2#3{Class.~and Quant.~Gr.~{\bf {#1}} (19{#2}) #3}
\def\cmp#1#2#3{Comm.~Math.~Phys.~{\bf {#1}} (19{#2}) #3}
\def\jmp#1#2#3{Jour.~Math.~Phys.~{\bf {#1}} (19{#2}) #3}
\def\ap#1#2#3{Ann.~of Phys.~{\bf {#1}} (19{#2}) #3}
\def\prep#1#2#3{Phys.~Rep.~{\bf {#1}C} (19{#2}) #3}
\def\ptp#1#2#3{Prog.~Theor.~Phys.~{\bf {#1}} (19{#2}) #3}
\def\ijmp#1#2#3{Int.~Jour.~Mod.~Phys.~{\bf {#1}} (19{#2}) #3}
\def\nc#1#2#3{Nuovo Cim.~{\bf {#1}} (19{#2}) #3}
\def\ibid#1#2#3{{\it ibid.}~{\bf {#1}} (19{#2}) #3}

\def\szet{{${\scriptstyle \b}$}}
\def\ula{{\un a}}
\def\ulb{{\un b}}
\def\ulc{{\un c}}
\def\uld{{\un d}}
\def\ulA{{\un A}}
\def\ulM{{\underline M}}
\def\cdm{{\Sc D}_{--}}
\def\cdp{{\Sc D}_{++}}
\def\vTheta{\check\Theta}
\def\Pisl{{\Pi\!\!\!\! /}}

\def\fracmm#1#2{{{#1}\over{#2}}}
\def\gg{{\hbox{\sc g}}}
\def\half{{\fracm12}}
\def\ha{\half}

\def\frac#1#2{{\textstyle{#1\over\vphantom2\smash{\raise -.20ex
        \hbox{$\scriptstyle{#2}$}}}}}                   

\def\fracm#1#2{\hbox{\large{${\frac{{#1}}{{#2}}}$}}}

\def\Dot#1{\buildrel{_{_{\hskip 0.01in}\bullet}}\over{#1}}
\def\dt#1{\Dot{#1}}
\def\uln{{\underline n}}
\def\Tilde#1{{\widetilde{#1}}\hskip 0.015in}
\def\Hat#1{\widehat{#1}}

\def\Dot#1{\buildrel{_{_{\hskip 0.01in}\bullet}}\over{#1}}
\def\dt#1{\Dot{#1}}
\def\gg{{\hbox{\sc g}}}
\def\nt{$~N=2$~}
\def\gg{{\hbox{\sc g}}}
\def\nt{$~N=2$~}
\def\tr{{\rm tr}}
\def\Tr{{\rm Tr}}
\def\mpl#1#2#3{Mod.~Phys.~Lett.~{\bf A{#1}} (19{#2}) #3}
\def\hati{{\hat i}} \def\hatj{{\hat j}} \def\hatk{{\hat k}}
\def\hatl{{\hat l}}

\oddsidemargin=0.03in
\evensidemargin=0.01in
\hsize=6.5in
\textwidth=6.5in

\noindent 1.~~{\it Introduction.~~~}The importance of self-dual
supersymmetric Yang-Mills (SDSYM) theory or self-dual supergravity
(SDSG) is based on the recent results in $~N=2$~ superstring theory
[1] that the consistent space-time background
is to be $~N=4$~ SDSYM for open $~N=2$~ superstring,
or $~N=8$~ SDSG for closed $~N=2$~ (heterotic) superstring
[2].  In particular, the $~N=4$~ maximal supersymmetry is needed for the
SDSYM, when the background superfield is irreducible [2].  As a
by-product, these developments have provided a solution to the
long-standing problem of invariant lagrangian for the SDSYM theory, by the
use of propagating multiplier fields necessarily appearing in $~N=2$~
superstring formulation [2].

        The mathematical conjecture [3] that all
possible bosonic integrable systems in lower-dimensions are coming from the
self-dual YM\footnotew{We use the abbreviation
YM for Yang-Mills, and SYM for supersymmetric YM.} theory in $~D=(2,2)$~
provides an additional
strong motivation for studying the self-dual theories.  Motivated by these
progresses both in physics and mathematics, we have recently presented a
series of SDSYM and SDSG theories in
$~2+2\-$~dimensions\footnotew{We use the symbol $~D=(t,s)$~ for the
space-time with the signature $~(+^t,-^s)$, {\it e.g.}, $~D=(1,3)$~ for
$~(+,-,-,-)$.  When the signature is not
very important, we also use the symbols like $~D=4$.} [4-8],
which are conjectured to generate lower-dimensional {\it supersymmetric}
integrable systems [9].
There are also other practical reasons
to study SDYM systems.  For example, we can
understand the connection between the inverse scattering method and the twister
methods through dimensional reduction (DR).  Exact
solutions to the $~D=2$~ integrable systems can also provide helpful clues
for the exact solutions for the original $~D=(2,2)$~ SDYM system.

        In this paper we take a one step further in the application of
the $~N=4$~ SDSYM, as the underlying ``master theory'' for
lower-dimensional models.  We show that a $~D=2$~ topological field
theory (TFT) [10]
with BRST-like symmetry proposed by Witten [11] is naturally generated by
an appropriate DR of $~N=2$~ SDSYM, which in turn is
obtained by a twisted truncation of the original $~N=4$~ SDSYM [2,7].
We also show how the Witten's TFT can accommodate the supersymmetric
Korteweg de Vries (SKdV) systems [12] or $~SL(n)$~ Toda field theory
[13], and study the large $~n$~ limit of the latter.

\bigskip\bigskip

\noindent 2.~~{\it Twisted Truncation of $~N=4$~ SDSYM into $~N=2$.~~~}We first
perform the truncation of the original $~N=4$~ SDSYM [2,7] into twisted $~N=2$~
theory, because the latter theory appears to be practically more useful
than the former.

        The field content of the $~N=4$~ SDSYM is $~(\hat A\du {\hat\m} I,
\hat{\Tilde\l}{}^I,\hat S\du\hati I,\hat T\du\hati I, \hat G\du{\hat\m\hat\n} I
,\hat\r^I)$~ [2,7], where we are using the similar notation as in
Ref\.[7].  The {\it hatted} fields and indices ({\it{}except for}
$~{\scst\hati,
{}~\hatj,~\cdots}$) are for $~D=(2,2)$~ to be distinguished from the
$~D=2$~ {\it unhatted} ones after our DR later.
This is the convention introduced in Ref\.[14].  The indices
$~{\scst I,~J,~\cdots}$~ are for the adjoint indices of
any arbitrary gauge group.\footnotew{We can choose any compact
as well as non-compact gauge groups such as $~SL(n)$,
as we will later.}~~The $~{\scst \hati,~\hatj,~\cdots~=~1,~2,~3}$~ are
related to the $~\a$~ and $\b\-$matrices described below.
The $~\hat G_{\m\n}{}^I$~ and $~\hat\r{}^I$~ are what we
call propagating multiplier fields, $~\hat A\du \m I $~ is the YM field,
$~\hat S\du \hati I$~ and $~\hat T\du \hati I$~ are scalars.  The
$~\hat{\Tilde\l}{}^I$~ is $~4\times 2\-$component {\it anti-chiral}
Majorana-Weyl
fermion [2,7], where the first $~4\-$components are for the $~N=4$~
supersymmetry, while the second $~2\-$components are for the {\it
anti-chiral} Majorana-Weyl spinor.  Relevantly, all the {\it tilded} (or
{\it untilded}) spinors are {\it anti-chiral} (or {\it chiral})
Majorana-Weyl spinors.  The
invariant lagrangian is [2,7]\footnotew{In this paper we are using
$~\g^\m$~ instead of $~\s^\ula$~ {\it etc.}~of Refs\.[4-8].
Accordingly, our $~\g$'s satisfy $~\{ \g_\m,\,\g_\n\} =
+ 2\eta_{\m\n}= 2 \hbox{diag.}~(+,+,-,-)$.
Needless to say, we also use the symbols like $~\g^{\m\n} \equiv
\g^{\[\m} \g^{\n\]} = (1/2) (\g^\m\g^\n - \g^\n\g^\m)$, where we do not put
any {\it tilde} for the first
$~\g$'s, following the convention in Ref\.[14].}
$$\eqalign{\hat \Lag{\,}_{D=4}^{N=4}= \,& - \fracm 12 \hat G^{\hat\m\hat\n\,I}
(\hat F\du{\hat\m\hat\n}I- \half\hat\e\du{\hat\m\hat\n}{\hat\r\hat\s}
\hat F\du{\hat\r\hat\s} I)
+ \fracm 12 (\hat D_{\hat\m} \hat S\du{\hat i} I)^2 - \fracm 12
(\hat D_{\hat\m}\hat
T\du{\hat i}I)^2 + 2i (\hat \r^I \hat\g^{\hat\m} \hat D_{\hat\m}
\hat{\Tilde\l}^I)  \cr
& - i f^{I J K} \left[ (\hat{\Tilde\l}{}^I \a_{\hat i}\hat{\Tilde\l}{}^J)
\hat S\du{\hat i} K + (\hat{\Tilde\l}{}^I \b_{\hat i} \hat{\Tilde\l}{}^J)
\hat T\du{\hat i} K \right] ~~, \cr }
\eqno(2.1) $$
where $~\hat D_{\hat\m}$~ is a gauge covariant derivative.
Notice the important role played by the multiplier fields $~\hat
G_{\hat\m\hat\n}$~
and $~\hat\r$, which made the lagrangian formulation possible.
This lagrangian is invariant under the $~N=4$~ supersymmetry [2,7]
$$\eqalign{&\d \hat A\du {\hat\m}I = -
i(\hat\e\hat\g_{\hat\m}\hat{\Tilde\l}{}^I) ~~, \cr
&\d \hat G_{\hat\m\hat\n} {}^I = 2i (\hat{\Tilde\e}\, \hat\g_{\[\hat\m}
\hat D _{\hat\n\]}\hat\r^I) -  i \fracm 12 f^{I J K} \hat\e \left(
\a_\hati S\du\hati J - \b_\hati T\du\hati J \right)
\hat\g_{\hat\m\hat\n} \hat\r^K ~~, \cr
&\d \hat \r^I = - \fracm 14 \hat\g^{\hat\m\hat\n} \hat\e \, \hat
G\du{\hat\m\hat\n} I
- \half \a_{\hat i} \hat\g^{\hat\m}\hat{\Tilde\e}\, \hat D_{\hat\m}\hat
S\du\hati I
- \half \b_{\hat i} \hat\g^{\hat\m}\hat{\Tilde\e} \, \hat D_{\hat\m}
\hat T_{\hat i}{}^I \cr
& ~~~~~ ~~~ + i \fracm 14 \hat\e^{\hat i\hat j\hat k} \a_\hati \hat\e\,
f^{I J K}
\hat S\du{\hat j} J \hat S\du \hatk K -i \fracm 14 \hat\e^{\hat i\hat j\hat k}
\b_{\hat i} \hat\e\, f^{I J K} \hat T\du{\hat j} J \hat T\du{\hat k} K
- \half f^{I J K} \a_{\hat j} \b_{\hat k} \hat\e\, \hat S\du{\hat j} J
\hat T\du{\hat k} K ~~, \cr
&\d \hat{\Tilde\l}^I = - \fracm 14 \hat\g^{\hat\m\hat\n}\,
\hat{\Tilde\e}\,
\hat F\du{\hat\m\hat\n} I -
\fracm 12 \a_{\hat i} \hat\g^{\hat\m}\hat\e\, \hat D_{\hat\m}\hat S
\du{\hat i}I + \half
\b_{\hat i} \hat\g^{\hat\m}\hat\e\, \hat D_{\hat\m}\hat T\du{\hat i}I ~~, \cr
&\d \hat S\du{\hat i} I = i (\hat\e\a_{\hat i} \hat \r^I) + i
(\hat{\Tilde\e}
\a_{\hat i} \hat{\Tilde\l}{}^I) ~~, ~~~~
\d \hat T\du{\hat i} I = i(\hat\e \b_{\hat i} \hat \r^I) - i
(\hat{\Tilde\e} \b_{\hat i} \hat{\Tilde\l}{}^I) ~~. \cr }
\eqno(2.2) $$
The matrices $~\a$~ and $~\b$~ $~4 \times 4 $~ matrices form the global $~SO(3)
\otimes SO(3)'$~ algebra:
$$ \[ \a _\hati, ~\a_\hatj \] = 2i \e_{\hati\hatj\hatk} \a_\hatk ~~,
{}~~~~\[ \b_\hati,~\b_\hatj\] = 2i \e_{\hati\hatj\hatk} \b_\hatk ~~,
{}~~~~\[\a_\hati,~\b_\hatj\] = 0 ~~.
\eqno(2.3) $$

        We are now ready for the twisted truncation of the $~N=4$~ to
get twisted $~N=2$~ SDSYM system.  First of all, we study the superspace
algebra of the former:
$$\eqalign{& \Big\{ \nabla_{\a A} , \Tilde\nabla_{\Dot\b B} \Big\} =
i (\s^\ula)_{\a \Dot\b} \d_{A B} \nabla_\ula ~~, \cr
& \{ \nabla_{\a A} , \nabla_{\b B} \} = - i C_{\a\b} (\a^\hati)_{A B} S\du
\hati I \t\low I +
i C_{\a\b} (\b^\hati)_{A B} T\du \hati I \t\low I ~~, \cr
& \Big\{ \Tilde\nabla_{\Dot\a A} , \Tilde\nabla_{\Dot\b B} \Big\} = -
i C_{\Dot\a\Dot\b} (\a^\hati)_{A B} S\du\hati I \t\low I - i C_{\Dot\a\Dot\b}
(\b^\hati)_{A B} T\du\hati I \t\low I ~~, \cr}
\eqno(2.4) $$
where we are temporarily using the superspace notation in Ref\.[7], and
$~\t\low I$~ are the generators of the YM gauge group.  The indices
$~{\scst A,~B,~\cdots}$~ are for the four-components for the $~N=4$~
acted on by the $~\a$~ and $~\b\-$matrices.  We recall that the original
global symmetry was $~SO(3) \otimes SO(3)'$, where the $~SO(3)$~ (or
$~SO(3)'$) corresponds to the $~\a$~ (or $~\b)\-$matrices above.  Our
twisting in the truncation into $~N=2$~ works as follows.  We first
identify $~SO(3)\otimes SO(3)' \approx Sp(1)\otimes Sp(1)'$.  Next we
realize only the diagonal subgroup $~SP(1)_D$~ of these two $~Sp(1)$~ groups,
and use the indices $~{\scst i,~j,~\cdots ~=~1,~2}$~ for the two-dimensional
fundamental representation of $~Sp(1)_D$, regarding the {\it undotted}
and {\it dotted} fermions as conjugate under the inner produces of
$~Sp(1)_D$.  To be more specific, for the {\it
undotted} $~{\scst \a,~\b,~\cdots}\-$indices we use the subscripts $~{\scst
{}_{i,~j,~\cdots}}$, while for the {\it dotted} indices $~{\scst
\Dot\a,~\Dot\b,~\cdots}$~ we use the superscripts $~^{\scst i,~j,~\cdots}$.
To be consistent with supersymmetry, we also need to
truncate the $~\hat S\du\hati I$~ and $~\hat T\du \hati I$~
{\it except for} $~\hat S\du 2
I\equiv S^I$~ and $~\hat T\du 2 I\equiv T^I$, and use $~i\s_2 = \e_{i j}$~
matrix for $~\a_2$~ and $~\b_2$, because they can be $~\a_2\equiv
\s_2 \otimes
I_2$~ and $~\b_2 \equiv I_2 \otimes \s_2$~ in some representation.
(Here $~\s_\hati$~ are the usual $~2\times 2$~ Pauli matrices.)
The resultant $~N=2$~ system has the global $~Sp(1)$~
symmetry whose algebra contains
$$\{ \nabla_{\a i }, \Tilde\nabla\du{\Dot\b} j \}  = i \d\du i j
(\s^\ula)_{\a\Dot \b} \nabla_\ula ~~.
\eqno(2.5) $$

Notice that the $~\d\du i j$~ on the r.h.s\.is $~Sp(1)\-$covariant,
i.e., the $~{\scst j}\-$index can be lowered by the metric tensor $~\e_
{i j}$, and eventually we get the ``twisted'' $~N=2$~ algebra
$$\eqalign{& \Big\{ \nabla_{\a i }, \Tilde\nabla_{\Dot\b j} \Big\}  =
i \e\low{i j} (\s^\ula)_{\a\Dot \b} \nabla_\ula ~~, \cr
&\{ \nabla_{\a i }, \nabla_{\b j} \}  = i C_{\a\b} \e\low{i j}
 \hat S^I\t\low I ~~, ~~~~
 \Big\{ \Tilde\nabla\du{\Dot\a}i , \Tilde\nabla\du{\Dot\b}j \Big\}  =
 i C_{\Dot\a\Dot\b} \e^{i j} \hat T^I\t\low I~~. \cr }
\eqno(2.6) $$
We are using the word ``twisted'', due to the vanishing {\it diagonal}
commutators with respect to the indices $~{\scst i}$~ and $~{\scst j}$.
All the resultant fermions have two-dimensional indices $~{\scst
i,~j,~\cdots~=~1,~2}$~ for $~N=2$.  The final field content of
our twisted $~N=2$~ is thus $~(\hat A\du{\hat\m} I, \hat G\du{\hat\m\hat\n} I,
\hat{\Tilde \l}\du i I, \hat \r\du i I, \hat S^I, \hat T^I )$.

        This truncation does not pose any problem about the
consistency of the system, because it is nothing else than putting
trivial (vanishing) classical solutions on the truncated fields.
In principle, we have
to check that all of these ``trivial'' solutions really satisfy all the
original field equations.  In practice, however,
we have to check the realization of
supersymmetry,\footnotew{This is true only at the classical level.  At
the quantum level we have to confirm anomaly cancellation
which is, however, harmless in the starting $~N=2$~ superstring
[2].} as will be confirmed to be $~N=2$~ below.  We
also mention that this sort of algebra, even though its twistedness was
not emphasized, has been already given in our previous paper [6] in the context
of non-lagrangian formulation.

        After this twisted truncation, we get the $~N=2$~ lagrangian
$$\eqalign{ \hat{\cal L}{\,}_{D=4}^{N=2}= \, &- \half \hat G^{\hat\m\hat\n \,
 I} \left( \hat F\du{\hat\m\hat\n} I -
 \half\hat\e\du{\hat\m\hat\n}{\hat\r\hat\s} \hat F\du{\hat\r\hat\s} I \right)
 - 2 (\hat D_{\hat\m} \hat S^I) (\hat D^{\hat \m} \hat T^I) \cr
& - 2i \big( \hat\r^{i I} \hat\g^{\hat\m} \hat D_{\hat\m} \hat{\Tilde\l} \du iI
 \big) - 2 f^{I J K} \big(\hat{\Tilde\l}{}^{i I} \hat{\Tilde\l} \du i J  \big)
 \hat S^K ~~, \cr }
\eqno(2.7)$$
invariant under the $~N=2$~ supersymmetry
$$\eqalign{& \d \hat A\du {\hat\m} I = - i (\hat\e^i \hat\g_{\hat\m}
 \hat{\Tilde\l} \du i I) ~~, \cr
&\d \hat G\du {\hat\m\hat\n} I = 2i \left( \hat{\Tilde\e}{}^i \hat\g_{\[
 \hat\m} \hat D_{\hat\n\]} \hat\r \du i I \right)
 - f^{I J K} \left(\hat\e^i \hat\g_{\hat\m\hat\n} \hat\r\du i J \right)
 \hat T^K ~~, \cr
&\d \hat\r \du i I = - \fracm1 4 \hat\g^{\hat\m\hat\n} \hat\e_i \hat
 G\du{\hat\m\hat\n} I + i \hat\g^{\hat\m} \hat{\Tilde\e}_i
 \hat D_{\hat\m} \hat S^I - f^{I J K} \hat S^J \hat T^K \hat \e_i~~, \cr
&\d \hat{\Tilde\l}\du i I = -\fracm 14 \hat\g^{\hat\m\hat\n}
 \hat{\Tilde\e}_i \hat F\du{\hat\m\hat\n} I
 - i \hat\g^{\hat\m} \hat{\Tilde\e}_i \hat D_{\hat\m} \hat T^I ~~, \cr
&\d \hat S^I = (\hat\e^i \hat\r\du i I) ~~, ~~~~
 \d \hat T^I = (\hat{\Tilde\e}{\,}^i \hat{\Tilde\l}\du i I ) ~~. \cr }
\eqno(2.8) $$
The invariance of (2.7) under (2.8) is easily confirmed.

We mention that this $~N=2$~ SDSYM lagrangian has many more applications
other than the DR we try in this paper, such as getting integrable
systems as in Ref\.[9].

\bigskip\bigskip

\noindent 3.~~{\it DR into $~D=2$.}~~~We have been so far
in the $~D=(2,2)$.  We can perform further a
DR to get the Witten's TFT in $~D=2$~ [11].  This TFT has a
peculiar symmetry, which resembles the nilpotent BRST symmetry
instead of the usual space-time supersymmetry.

        There are some ways of DR to realize such nilpotency.  One way is
to truncate one of the two
supersymmetries in the $~N=2$.  Recall our twisted $~N=2$~
supersymmetry algebra (2.6).  Due to the existence of $~\e_{i j}$,
it is now straightforward to recognize that the
nilpotency is realized by truncating one of the $~\nabla_{\a i}$~ and
both of the $~\Tilde\nabla_{\Dot\a i}$.  Eventually all the r.h.s\.in (2.6)
vanish everywhere, and we get the nilpotency $~\{
\nabla_{\a 1}, \nabla_{\b 1}\} = 0$.
In terms of the component transformation
rules (2.8), this amounts to the requirements
$$\hat{\Tilde\e} _i = 0~~, ~~~~ \hat S^I = \hat T^I = 0~~,
{}~~~~\hat {\Tilde\l}\du 2I = 0
{}~~, ~~~~\hat \r\du 1 I = 0 ~~, ~~~~\hat\e_1 = 0~~.
\eqno(3.1) $$

        We also need some arrangements for the gamma-matrices for our
DR.  In $~D=(2,2)$~ originally we had [4,7]
$$\hat\g^{\hat\m} = (\hat\g^1,\,\hat\g^2,\,\hat\g^3,\,\hat\g^4)
= (- i\s_1,\, - i \s_2, \, - \s_3, \, i I_2) ~~,
\eqno(3.2) $$
when $~\hat\g^{\hat\m}$~ is acting on two-component
{\it anti-chiral} Majorana Weyl
spinors,\footnotew{When acting on {\it chiral} Majorana-Weyl spinors
instead, only $~\hat\g^1,~\hat\g^2$~ and $~\hat\g^3$~ change the signs, {\it
e.g.}, $~\hat\g^1= +i \s_1$.  (See eq\.(B.8) in Ref\.[7].)}
and $~\s_1,~\s_2,~\s_3$ are the usual $~2\times 2$~ Pauli
matrices.  After the DR, we get
$$\hat\g^\m = \g^\m = - i \e^{\m\n} \g_5 \g_\n ~~, ~~~~ {\scst (\m,~\n,~
\cdots ~=~ 3,~4)} ~~,
\eqno(3.3) $$
where $~\g_5 \equiv - i\hat\g^3\hat\g^4 = \s_3$~ is for the chiral
projection in $~D=2$, and $~\g^\m$'s~ with {\it no hats} are the
$~\g\-$matrices in $~D=2$~.  The reason of choosing $~{\scst \m,~\n,~=~3,~4}$~
as the $~D=2$~ coordinates instead of $~{\scst \m,~\n,~\cdots~=~ 1,~2}$~ is
that $~\g^3$~ and $~\g^4$~ are diagonal in
the representation (3.2), which is convenient for chiral projections.
Accordingly, the final space-time is Euclidean with
the signature $~D=(0,2)$.  Additionally we need the usual simple
DR rules and other appropriate truncations:
$$\eqalign{ &\partial_\a = 0 ~~, ~~~~ \hat A\du\a I = 0 ~~, ~~~~\hat A\du \m I
= A\du \m I ~~, \cr
& \hat G_{\a\b}{}^I = 0 ~~, ~~~~\hat G\du{\m \a} I = 0 ~~, ~~~~{\scst
(\a,~\b,~\cdots ~=~ 1,~2)}~~. \cr}
\eqno(3.4) $$
Here the coordinates $~{\scst\a,~\b.~\cdots ~=~1,~2}$~ are for the ``extra''
coordinates, which are truncated in the simple DR.

        Applying our DR rules (3.1) - (3.4) to (2.7), and after
introducing a new scalar field $~\phi^I \equiv - (1/4) \e^{\m\n} G\du {\m\n}
I$~ and a vector spinor $~\psi\du\m I
\equiv - i \g_5\g_\m\Tilde\l\du 1I$,\footnotew{The latter replacement poses
{\it no} problem, even though it is not generally invertible.
This is because the field $~\Tilde\l$~ appears everywhere only in the
combination of $~i\g_5\g_\m \Tilde\l$, so that the $~\psi_\m$~ is
defined uniquely
up to ``gauge'' transformation like $~\d\psi_\m = \left(\d\du \m \n-
\g\du\m\n\right) \e_\n~$ with a parameter $~\e_\n$.} together with
$~\r^I\equiv \r\du 2 I$, we get the $~D=2$~
lagrangian
$${\cal L}{\,}^{\rm{TFT}}_{D=2} = \, \e^{\m\n} \phi^I
F\du{\m\n} I - i \e^{\m\n} (\r ^I T\du{\m\n} I ) ~~,
\eqno(3.5)$$
with $~T\du{\m\n}I \equiv D_\m \psi\du \n I - D_\n \psi\du\m I $,
which is nothing else than the Witten's TFT.  The fermions
$~\psi\du \m I$~ and $~\r^I$~
are Majorana-Weyl, satisfying $~\g_5 \psi_\m{}^I = \psi_\m {}^I,
{}~\g_5 \r^I = \r^I$.
This lagrangian is an extension of the $~\phi F\-$type lagrangian
studied in Ref\.[15].  We can also confirm that the BRST-like symmetry of
(3.5) also arises after the DR of (2.8),
after the identification $~\e^1 = \e_2 \equiv \e= \g_5 \e$,
$$\eqalign{&\d A\du \m I = i (\e \psi\du\m I) ~~, ~~~~ \d\phi^I = 0 ~~,
\cr
&\d\r^I = \e \phi^I~~, ~~~~ \d\psi\du \m I = 0~~. \cr}
\eqno(3.6) $$
As is well-known, the $~\phi^I\-$equation out of (3.5) gives
$~F_{\m\n}{}^I = 0~$ which can embed innumerable (bosonic) integrable
systems in $~D=2$, as has been analyzed in Ref\.[16].

\bigskip\bigskip

\noindent 4.~~{\it Embedding of SKdV System.}~~~As a simplest
application of the Witten's TFT we re-obtained, we show how the $~N=1$~
SKdV system [12] can be embedded, when the gauge group is
$~U(1)$.  In this case the field equations of
$~\phi$~ and $~\r$~ give the vanishing field strengths:
$$\eqalign{&F_{x t } = \partial_x A_t - \partial _t A_x = 0 ~~, \cr
& T_{x t} = \partial_x \psi_t - \partial_t \psi _x = 0 ~~. \cr }
\eqno(4.1) $$
Now our embedding is
$$\eqalign{&A_x = u ~~, ~~~~A_t = - u '' + 3u^2 - 3\xi \xi ' ~~, \cr
& \psi_x = \xi ~~, ~~~~ \psi_t = - \xi'' + 3 u \xi ~~. \cr}
\eqno(4.2) $$
Insertion of this into (4.1) yields the standard $~N=1$~ SKdV equations
[12]:
$$\Dot u = - u''' + 6 u u' - 3\xi\xi ''~~, ~~~~
\Dot \xi = - \xi ''' + 3 (u\xi )' ~~.
\eqno(4.3)$$
This $~U(1)\-$embedding is similar to our previous results [9].

        At the lagrangian level (3.4) we get
$$\Lag_{\rm SKdV} = -2 \phi \left( \Dot u + u''' - 6 u u' + 3\xi\xi'' \right)
+2 i \r \left(\Dot\xi+ \xi''' - 3u\xi' - 3 u' \xi \right) ~~.
\eqno(4.4) $$
This is invariant under the $~N=1$~ supersymmetry
$$\eqalign{& \d u = \e \xi' ~~, ~~~~ \d\xi = \e u ~~, \cr
& \d\phi = -i \e\r ~~, ~~~~ \d\r = i \e \phi' ~~. \cr}
\eqno(4.5) $$
Interestingly, the $~N=1$~ {\it supersymmetry} manifests itself
as a hidden symmetry out of a system originally only with the BRST-like
symmetry.

\bigskip\bigskip

\noindent 5.~~{\it $~W_\infty$~ as $~{n\rightarrow\infty}$~ Limit of
$~SL(n)\-$Toda Theory.}~~~We next show how $~SL(n)\-$Toda theory can be
embedded into the bosonic part [15] of the TFT, and take the
$~n\rightarrow \infty$~ limit to get $~W_\infty\-$algebra [17].  This is
similar to the approach in Ref\.[18], where such a limit was taken at the
field equation level of Wess-Zumino-Novikov-Witten model, while we can
deal with a lagrangian of TFT.  In this section, we look only into the
purely bosonic $~\phi F\-$part of the total lagrangian (3.4).

        As usual we can embed the $~SL(n)\-$Toda theory into the
on-shell vanishing field strength, as
$$A_z = - \partial_z \Tilde\varphi \cdot H + m \sum_{i=1}^n e_i ^+ ~~, ~~~~
A_{\Bar z} = m \sum_{i=1}^n  e^{-\a_i \cdot \Tilde \varphi} e _i ^ -
{}~~,
\eqno(5.1) $$
with a constant $~m$.
For convenience sake, we use the $~D=2$~ complex coordinates $~z\equiv
x+i t,~\Bar z\equiv x-i t$, and the Cartan-Weyl bases $~(H,e_i^+,e_i^-)~{\scst
(i,~j,~\cdots~=~1,~\cdots,~n)}$, satisfying
$$\[ H,H \] = 0 ~~, ~~~~ \[ H, e_i^\pm \] = \pm \a_i e_i ^\pm ~~, ~~~~
\[ e_i^+, e_j ^- \] = \d\low{i j} \a_i \cdot H~~.
\eqno(5.2) $$
The $~\a_i$'s are the simple roots regarded as $~r\-$dimensional
vectors, and $~H$~ is another $~r\-$vector in the Cartan subalgebra,
such that $~h_i = \a_i \cdot H, ~ \[ h_i,h_j\] = 0 , ~ \[ h_i,e_j^\pm \]
= \pm a_{i j} e_j^\pm, ~ \[ e_i ^+ , e _j ^- \] = \d_{ i j } h_i $,
with the Cartan-Weyl matrix $~a_{i j} = \a_i \cdot \a_j$.  Eventually
$~F_{z\Bar z}$~ is
$$F_{z \Bar z} = \sum_i \left[ \partial_z\partial_{\Bar z}
\Tilde\varphi _i + m^2\exp\left(- \sum_j a_{i j} \Tilde \varphi _j \right)
\right] h_i ~~,
\eqno(5.3) $$
for $~\Tilde\varphi = \sum _i \Tilde\varphi_i \a_i $.
To accord with the standard notation in the
$~SL(n)\-$Toda theory, we use the orthonormal basses $~e_i$~ such that
$~\a_i = e_i - e_{i+1}$.  (See, {\it e.g.}, Ref.~[19].)
Using the expansion $~\Tilde \varphi = \sum _i \varphi_i
e_i$, so that $~\varphi _i = \Tilde\varphi_i - \Tilde\varphi_{i-1}$~
together with $~\phi = \sum _i \phi_i e_i$, we get
$$\Lag_{SL(n)} = \Lag_{\phi F} =
\sum_i \phi_i \left[ \partial_z\partial_{\Bar z} \varphi_i + m^2
\left(e^{\varphi_{i+1} - \varphi_i} - e^{\varphi_i - \varphi_{i-1}}
\right)\right]~~.
\eqno(5.4) $$
Here it is understood that $~\varphi_0 = 0,~\varphi_{n+1}=0$.  This
yields the $~SL(n)\-$Toda equation [13] out of multiplier fields $~\phi_i$.

        It is important to notice that the symmetry transformation for
the Toda theory is subject to a certain constraint.
This constraint can be obtained by requiring that the {\it on-shell}
invariance of the field strength $~F_{z\Bar z} =
0$~ in (5.3) is maintained after infinitesimal gauge transformation
$~\d\varphi_i$.  The constraint thus obtained is
$$\partial_z\partial_{\Bar z} (\d\varphi_i) + m^2 \d \left[ e^{\varphi_{i+1}
- \varphi_i} - e ^{\varphi_i - \varphi_{i-1}} \right] = 0 ~~.
\eqno(5.5) $$
This symmetry contains what is called
extended conformal symmetry of Zamolodchikov's
$~W_n \-$algebra [20,21], including spin $~k\-$transformations.  For our later
purpose it is sufficient to consider the spin $~2\-$transformation [18,20]:
$$\d(\varphi_i - \varphi_{i-1}) = \partial_{\Bar z} \e(\Bar z) + \e(\Bar
z) \partial_{\Bar z} (\varphi_i - \varphi_{i-1})~~,
\eqno(5.6) $$
where $~\e (\Bar z)$~ is an arbitrary anti-holomorphic infinitesimal
function.  It is not difficult to show that (5.6) satisfies the
constraint (5.5).  Relevantly, the $~\phi_i\-$field transforms
$$\d(\phi_i- \phi_{i-1}) = \e(\Bar z) \partial_{\Bar z} (\phi_i -
\phi_{i-1})  ~~.
\eqno(5.7) $$

        We now take the $~n\rightarrow\infty$~ limit to see if the
$~W_\infty\-$algebra can be implemented.  This process is essentially
the same as in Ref\.[18].  We introduce a new coordinate $~\t$~ and
new continuous functions $~v(x,t,\t)$~ and $~\Tilde\L(x,t,\t)$~ in
three-dimensions such that
$$\varphi_i(x,t) = \fracmm1\Delta v(x,t,\t= \t_0 +i\Delta) ~~,
{}~~~~\phi_i(x,t) = \fracmm1\Delta \Tilde\L (x,t,\t=\t_0+i\Delta) ~~,
\eqno(5.8) $$
where $~\Delta\equiv (\t_f-\t_0)/n$~ for the domain $~\t_0 \le \t \le
\t_f$, and we require also $~m \equiv 1/\Delta$.
Simple algebra reveals that $~\varphi_{i+1} - \varphi_i$~
corresponds to a derivative $~\partial_\t v$~ after the limit $~n\rightarrow
\infty$, and the $~i\-$summation becomes nothing else than a
$~\t\-$integration, if we supply an overall factor $~\Delta^3$.  In
fact, for arbitrary functions $~f(\t)$~ and $~g(\t)$~ such that
$$f_i \equiv \fracmm 1\Delta f(\t=\t_0 + i\Delta) ~~, ~~~~ g_i = \fracmm
1\Delta g(\t= \t_0 + i\Delta) ~~,
\eqno(5.9) $$
we get
$$\Delta^3 \sum_i f_i g_i = \Delta^3 \fracmm
1{\Delta^2} \sum _i f(\t + i \Delta) \, g(\t + i \Delta) ~~
{\buildrel{\scst n\rightarrow \infty} \over \longrightarrow} ~~
\int_{\t_0} ^{\t_f} d\t \, f(\t) \, g(\t) ~~.
\eqno(5.10) $$
After all, the $~\phi F\-$action in (5.4) tends to
$$\eqalign{I_{SL(n)}~~ {\buildrel{\scst n\rightarrow
\infty} \over \longrightarrow}~~
I_\infty = & \int d x \int d t \int_{\t_0} ^{\t_f} d\t\, \Tilde\L \left[
\partial_z\partial_{\Bar z} v + \partial_\t e^{(\partial_\t v)} \right]
\cr
= & \int d x \int d t \int_{\t_0}^{\t_f} d\t \, \L
\left[ \partial_z\partial_{\Bar z} u + \partial_\t^2(e^u) \right]  ~~, \cr}
\eqno(5.11) $$
where $~\Tilde\L\equiv - \partial_\t \L$~ and $~u \equiv \partial_\t v$~
after partial integration for $~\t$.  This action (5.11) yields nothing
else than Pleba{\~ n}ski's first heavenly equation with one rotational
Killing symmetry [22]:
$$\partial_z\partial_{\Bar z} u + \partial_\t^2 (e^u) = 0 ~~,
\eqno(5.12) $$
which is subject to the sub-algebra of surface
area-preserving diffeomorphism symmetry $~\hbox{\it sdiff}_+(R^2)$~
as the $~n\rightarrow\infty$~
limit [17] of $~W_n\-$symmetry [20] inherent in the $~SL(n)\-$Toda
theory.
This corresponds to the invariance of $~I_\infty
$~ (5.11) under the large $n\-$limit of the transformations (5.6) and
(5.7), namely
$$\d u = \partial_{\Bar z} \e + \e \partial_{\Bar z} u ~~, ~~~~
\d\L = \e \partial_{\Bar z} \L~~.
\eqno(5.13) $$
The infinitesimal parameter $~\e$~ satisfies the large $~n\-$limit of
the constraint (5.5):\footnotew{The calculation
is so far based on the anti-holomorphicity of $~\e$, and $~u$~ is a complex
function, but from now on this restriction is released.
This is because we can repeat the
same procedure for the holomorphic sector, and combine it with the
anti-holomorphic sector to make $~u$~ real [18].}
$$\partial_z\partial_{\Bar z} \e + (\partial_z\e)(\partial_{\Bar z} u )
= - 2 (\partial_\t e^u)(\partial_\t \e) - e^u \partial_\t^2 \e ~~.
\eqno(5.14) $$
Eq.~(5.13) is nothing but the usual area-preserving diffeomorphism for
a scalar {\it density} $~u$~
and a scalar $~\L$.
Thus we conclude that the (bosonic part of) TFT (3.4) really
accommodate the $~W_\infty\-$gravity as the large $~n\-$limit, {\it via}
$~SL(n)\-$Toda field theory.

\bigskip\bigskip

\noindent 6.~~{\it Concluding Remarks.}~~~In this paper we have shown
that the Witten's TFT will come out of a DR and a twisted truncation of
the $~N=4$~ SDSYM which is the consistent background of the
$~N=2$~ open superstring [1,2].  It had been already well-known that such a
TFT is obtained from the $~N=2$~ SYM in $~D=4$.  However,
the important
ingredient in our results is the fact that the same theory can come out even
from the SDSYM theory.  As an important application of the Witten's TFT,
we gave examples of embedding the $~N=1$~ SKdV systems and the
$~SL(n)\-$Toda theory into its purely bosonic part,
and we further studied the large $~n\-$limit to see how
$~W_\infty\-$gravity comes out.

        Our DR applied to a twisted $~N=2$~ SDSYM has generated a TFT
with the peculiar BRST-like symmetry in lower-dimensions.  In
particular, the twisting has played an important role truncating one
of the two supersymmetries of $~N=2$, as characterized by the {\it
non-diagonal} $~\e_{i j}\-$tensors in (2.6).
The idea of twisting was first introduced in $~D=2~$ with $~N=2$~
supersymmetry [23-25] out of a DR of $~D=4,\, N=2$~ supersymmetry.
It is therefore a next natural step to seek a ``twisting'' of
DR applied to $~D=4$~ SDSYM system that can generate other twisted TFT in
$~D=2$.

To our knowledge, our trial is the first one getting a TFT in
$~D=2$~ by a DR applied to the SDSYM theory.
Our results in this paper elucidate how rich the original $~N=4$~ SDSYM theory
is embedding the $~D=(0,2)$~ TFT which is itself
a master theory of (topological) $~W_n\-$gravity, matrix models, and
supersymmetric KdV systems in $~D=2$.  Therefore other twisted theories such
as in Ref\.[24] may also have BRST-like symmetry in $~D=4$.
However, these theories are unlikely to be related to $~N=2$~
superstring, if the analysis in Ref\.[2] is correct about the uniqueness of
its consistent background.

        Our result in this paper has established a clear link between
TFT with twisted BRST-like symmetry and $~N=2$~ superstring,
{\it via} $~N=4$~ SDSYM as its consistent background.  This further
suggests that the $~N=2$~ superstring is really the underlying
theory of other superstring theories, which are based on $~D=2$~
superconformal
theories.  To put it differently, the idea of ``world-sheet for
world-sheet'' or ``string for string'' presented in Refs\.[26,27] is not
illusional, but is supported by the solid mathematical background
elucidated by the TFT.

        Even though we did not show in this paper the embeddings of
other interesting models such as the $~(n-1)\-$matrix models,
$~(n-1)\-$generlized KdV hierarchy, $~SL(2)$~ topological gravity, {\it
etc}., these are easily shown to be generated by putting appropriate
ans{\" a}tze.  The advantage of the lagrangian formulation is to make all
the symmetries crystallized in the action invariance.
The matrix models are originally developed in the context of
non-critical string, which are further shown to be different
manifestation of $~W_n\-$gravity [28].  Our result provides another
viewpoint that these lower-dimensional systems are directly related to,
or even the descendants of the $~N=2$~ superstring after appropriate
twisted truncation and DR procedures.

        In our recent paper [29], we have shown how the supersymmetric
KP systems in $~D=2$~ can be embedded into the SDSYM.  If the TFT
treated in this paper governs the $~D=2$~ integrable systems, we expect
corresponding TFT (resembling Chern-Simons theory) in $~D=3$.

\bigskip\bigskip

We are grateful to E\.Witten for explaining the significance of
twisted truncation and dimensional reductions.  We are also indebted
to S.J\.Gates, Jr.~for valuable suggestions.

\bigskip\bigskip

\vfill\eject

\refs
\small

\items{1} H.~Ooguri and C.~Vafa, \mpl{5}{90}{1389};
\np{361}{91}{469}; \ibid{367}{91}{83};
\item{  } H.~Nishino and S.J.~Gates, Jr., \mpl{7}{92}{2543}.

\items{2} W.~Siegel, Stony Brook preprint, ITP-SB-92-24 (May, 1992).

\items{3} M.F.~Atiyah, unpublished;
\item{  } R.S\.Ward, Phil.~Trans.~Roy.~Lond.~{\bf A315} (1985) 451;
\item{  } N.J\.Hitchin, Proc.~Lond.~Math.~Soc.~{\bf 55} (1987) 59.

\items{4} S.V.~Ketov, S.J.~Gates and H.~Nishino, Maryland preprint,
UMDEPP 92--163.

\items{5} H.~Nishino, S.J.~Gates, Jr. and S.V.~Ketov,
Maryland preprint, UMDEPP 92--171.

\items{6} S.J.~Gates, Jr., H.~Nishino and S.V.~Ketov,
\pl{297}{92}{99}.

\items{7} S.J.~Ketov, H.~Nishino and S.J.~Gates, Jr., Maryland preprint,
UMDEPP 92--211 (June 1992), to appear in Nucl.~Phys.~B.

\items{8} H.~Nishino, Maryland preprint, UMDEPP 93-79.

\items{9} S.J.~Gates, Jr.~and H.~Nishino, \pl{299}{93}{255}.

\items{10} For reviews, see {\it e.g.}, D.~Birmingham, M.~Blau, M.~Rakowski and
G.~Thompson, \prep{209}{91}{129}.

\items{11} E.~Witten, in Proceedings of String 1990,
College Station Workshop, p.$\,$50.

\items{12} Yu.I.~Manin and \& A.O.~Radul, \cmp{98}{85}{65};
\item{  } P.~Mathieu, Jour.~Math.~Phys.~{\bf 29} (1988) 2499;
\item{  } C.A.~Laberge and P.~Mathieu, \pl{215}{88}{718};
\item{  } T.~Inami and H.~Kanno, \ijmp{A7}{92}{419}.

\items{13} See, {\it e.g.}, P.~Mansfield, \np{208}{82}{277}; and references
therein.

\items{14} H.~Nishino, Maryland preprint, UMDEPP 93-52.

\items{15} A.~Chamseddine and D.~Wyler, \pl{228}{89}{75}; \np{340}{90}{595}.

\items{16} See, {\it e.g.}, T.A.~Ivanova and A.D.~Popov, \pla{170}{92}{293}.

\items{17} See, {\it e.g.}, E.~Bergshoeff, C.N.~Pope, L.J.~Romans,
E.~Sezgin, X.~Shen and K.~Stelle, \pl{243}{90}{350};
\item{  } E.~Sezgin, Texas A \& M preprint,
CTP-TAMU-13/92 (Feb.~1991); and references therein;
\item{  } C.M.~Hull, Queen Mary preprint, QMW-93-2;
\item{  } I.~Bakas, \pl{228}{89}{57}.

\items{18} Q.H.~Park, \pl{236}{90}{429}; \np{333}{90}{267}.

\items{19} V.~Kac, Adv.~Math.~{\bf 26} (1978) 8.

\items{20} A.~Bilal and J.L.~Gervais, \pl{206}{88}{412}.

\items{21} A.B.~Zamolodchikov, Theor.~Math.~Phys.~{\bf 99} (1985) 108;
\item{  } V.~Fateev and S.~Lukyanov, \ijmp{3}{88}{507}.

\items{22} J.F.~Pleba{\~ n}ski, \jmp{23}{82}{1126};
\item{  } J.~Gegenberg and A.~Das, \grg{16}{84}{817}.

\items{23} S.J.~Gates, \np{238}{84}{349};
\item{  } S.J.~Gates, Jr., C.M.~Hull and M.~Ro{\v c}ek, \np{248}{84}{157}.

\items{24} T.~Eguchi and S.K.~Yang, \mpl{5}{90}{1693};
\item{  } E.~Witten, \cmp{118}{88}411{}; \np{371}{92}{191}.

\items{25} E.~Witten, \cmp{117}{88}{353}.

\items{26} M.~Green, \pl{193}{87}{439}; \np{293}{87}{593}.

\items{27} H.~Nishino, \mpl{7}{92}{1805};
\item{ } H.~Nishino and S.J.~Gates, Maryland preprint, UMDEPP 92-060,
to appear in
\item{  } Int.~Jour.~Mod.~Phys.

\items{28} See {\it e.g.}, M.F.~Awada and Z.~Qiu, \pl{245}{90}{85} and 359.

\items{29} H.~Nishino, Maryland preprint, UMDEPP 93-145 (Feb.~1993).

\end{document}